\documentclass[11pt,a4paper]{article}
\pdfoutput=1
\usepackage[T1]{fontenc}
\usepackage{jcappub}
\usepackage{microtype}

\newcommand{\be}{\begin{equation}}
\newcommand{\ee}{\end{equation}}
\newcommand{\bea}{\begin{eqnarray}}
\newcommand{\ena}{\end{eqnarray}}

\newcommand\m{\ensuremath{\mu}}

\newcommand{\de}{\partial}

\newcommand{\ba}{\begin{eqnarray}}
\newcommand{\ea}{\end{eqnarray}}
\newcommand{\plm}{M_{\text{Pl}}}

\title{\boldmath Resilience of long modes in cosmological observables}

\author[a,b,c,d]{Sabino Matarrese,}
\author[e,f]{Luigi Pilo}
\author[d,f]{and Rocco Rollo}

\affiliation[a]{Dipartimento di Fisica e Astronomia "Galileo Galilei",
Universit\`{a} degli Studi di Padova,\\
 via Marzolo 8, I-35131 Padova,Italy}
\affiliation[b]{INFN, Sezione di Padova,\\
 via Marzolo 8, I-35131 Padova, Italy}
\affiliation[c]{INAF-Osservatorio Astronomico di Padova,\\
Vicolo dell' Osservatorio 5, I-35122 Padova, Italy}
\affiliation[d]{Gran Sasso Science Institute (GSSI),\\
Viale Francesco Crispi 7, I-67100 L'Aquila, Italy}
\affiliation[e]{Dipartimento di Ingegneria e Scienze dell'Informazione
  e Matematica, Universit\`a degli Studi dell'Aquila,\\
Via Vetoio (Coppito 1),I-67010 L'Aquila, Italy}
\affiliation[f]{INFN, Laboratori Nazionali del Gran Sasso,\\
I-67010 Assergi, Italy}

\emailAdd{sabino.matarrese@pd.infn.it}
\emailAdd{luigi.pilo@aquila.infn.it}
\emailAdd{rocco.rollo@gssi.it}

\abstract{
  By a careful implementation of gauge transformations involving long-wavelength modes, we show that a variety of effects involving 
squeezed bispectrum configurations, for which one Fourier mode is much shorter than the other two, {\it cannot} be gauged away, except for the unphysical {\it exactly} infinite-wavelength ($k=0$) limit. Our result applies, in particular, to the Maldacena consistency relation for single-field inflation, yielding a {\it local} non-Gaussianity strength 
$f_{\rm NL}^{\rm local} = - (5/12)(n_S-1)$ (with $n_S$ the primordial spectral index of scalar perturbations), and to the $f_{\rm NL}^{\rm GR} = -5/3$ term, appearing in the 
dark matter bispectrum and in the halo bias, as a consequence of the general relativistic non-linear evolution of matter perturbations. Such effects are therefore physical and observable in principle by future high-sensitivity experiments.
}


\arxivnumber{1812.03844}

\begin{document}
\maketitle
\flushbottom

\section{Introduction}

The fundamental role of long-wavelength perturbations in cosmology has been recognised in various contexts: the study of primordial non-Gaussianities, the analysis of non-linearities 
in the dark matter and halo clustering properties, the study of specific secondary anisotropies in the Cosmic Microwave Background (CMB). In all these cases, the role of long-wavelength modes can be accounted for the calculation of specific observables (such as the dark matter bispectrum, the halo bias or the CMB bispectrum) by either adopting traditional higher-order perturbation theory 
techniques, or by considering - as a sort of useful short-cut tool -
long modes as a local background (a ``Separate Universe") on top of
which the contribution by shorter-wavelength perturbations can
accounted for, by means of a linear analysis, though the two
techniques are completely equivalent
The strength of non-Gaussianity $f_{\rm NL}$, which in the simplest single-field ``standard inflation"
case is  of the order of the first slow-roll parameters
$\epsilon$ and $\eta$ \cite{Gangui:1993tt,Gangui:1999vg,Wang:1999vf,Acquaviva:2002ud,Maldacena:2002vr,Lyth:2005du}, contains a {\it
  local|} contribution, whose dominant signal resides in {\it
  squeezed} bispectrum triangles (i.e. such that one triangle side in
Fourier space is much shorter than the other two), given by
\cite{Maldacena:2002vr} $f_{\rm NL}^{\rm local} = - (5/12)(n_S-1)$, where $n_S$ is the primordial spectral index of scalar
perturbations. 
The leading contribution to the bispectrum in the squeezed limit, for single-field standard inflation models, can be  predicted
without actually calculating the 3-point function but relying only on
the 2-point one
\cite{Creminelli:2004pv,Cheung:2007sv,Creminelli:2011rh,Senatore:2012wy};
the so-called ``Maldacena consistency relation'' connects the bispectrum and the power-spectrum. The consistency condition in
single-field inflation can also be derived by exploiting the residual
symmetries of the gauge-fixed action for the comoving curvature
perturbation $\zeta$ \cite{Creminelli:2012ed,Hinterbichler:2012nm,Hinterbichler:2013dpa,Hui:2018cag}.
This prediction has to be compared with current upper bounds derived
by the analysis of {\it Planck} satellite CMB data ($f_{\rm NL}^{\rm
  local} = - 0.9 \pm 5.1$, \cite{Akrami:2019izv}), while for the
spectral tilt one has $n_S=
 0.9652 \pm 0.0042$  \cite{Aghanim:2018eyx}, thus making this theoretical prediction still far from detectability. Observational prospects on CMB temperature and polarisation data and new large-scale structure surveys will have the capability to tighten the primordial non-Gaussianity bound to $f_{\rm NL}^{\rm local} = {\cal O}(1)$  \cite{Finelli:2016cyd,Carbone:2008iz,Dore:2014cca,Karagiannis:2018jdt}. More futuristic projects on CMB spectral distortions \cite{Pajer:2012vz} and measurements of 21cm background fluctuations from the dark ages (see, e.g. \cite{Munoz:2015eqa} and refs. therein) may eventually reach the required sensitivity $f_{\rm NL}^{\rm local}  \sim {\cal O}(10^{-2})$.

The long-short splitting technique was later adopted to study the second-order Sachs-Wolfe contribution to CMB anisotropies \cite{Bartolo:2005fp,Boubekeur:2009uk}, whose calculation confirms previous findings obtained by standard second-order perturbative techniques \cite{Mollerach:1997up,Bartolo:2004ty}. Similarly, the CMB non-Gaussianity arising from non-linear effects at recombination was successfully obtained by both techniques \cite{Bartolo:2011wb,Creminelli:2011sq}. 

As soon as general relativistic calculations of the dark matter density perturbations were performed up to second-order (see, e.g. \cite{Bartolo:2005xa}), a term mimicking local non-Gaussianity was found, as a manifestation of long-wavelength gravitational potential modes, which, once inside the horizon, modulate short-wavelength density perturbations. The amplitude of this ``General Relativistic" 
non-linear effect (dubbed GR-non-Gaussianity, because of its genuinely non-Newtonian origin) is equivalent to a local non-Gaussianity strength $f_{\rm NL}^{\rm GR} = - 5/3$ in the matter bispectrum. Later calculations confirmed that result and extended it to all perturbative orders \cite{Bruni:2014xma}.  
The same term was then studied in connection with local non-Gaussianity corrections to the dark matter halo bias, where it manifests itself as a scale-dependent bias contribution, thereby adding to any primordial $f_{\rm NL}^{\rm local}$-like term \cite{Dalal:2007cu,Matarrese:2008nc}. This would then make such a GR effect observable in the galaxy power-spectrum \cite{Verde:2009hy} obtained from ongoing galaxy survey data.

An important issue however arises in this framework and has been long
debated in the literature \cite{Bartolo:2005xa,Verde:2009hy,
  Villa_2014, Bruni:2014xma, Camera:2014bwa,
  Creminelli:2011rh,Creminelli:2011sq,Tanaka_2011, dePutter:2015vga,
  Baldauf_2011, Horn_2015, Pajer:2012vz, Pajer:2013ana, Dai:2015jaa,
  Dai:2015rda, Bartolo:2015qva,Cabass:2016cgp, Cabass:2018roz,
  Koyama:2018ttg, Umeh:2019qyd}, especially in connection with the
observability of the consistency relation and of the GR contribution
to halo bias: do these terms correspond to a physical observable
effect, or can they be cancelled by a suitable coordinates
tranformation (spatial dilatation)? 
This cancellation has been often claimed as a manifestation of the
Equivalence Principle \cite{Grimm:2020ays}, according to which the acceleration caused by a
{\it uniform} gravitational field cannot be locally distinguished from that caused by a non-inertial reference frame.

The aim of this paper is to provide a critical discussion of the
claim that the squeezed limit of ``single-clock" primordial non-Gaussianity is not observable 
\cite{Pajer:2013ana,Tanaka_2011,Pajer:2013ana,
  Cabass:2016cgp,Bravo:2017gct}. 
We find that long perturbation modes, containing any wavenumber $k\neq
0$, {\it cannot be gauged away}, neither by a spatial dilatation
\cite{dePutter:2015vga, Baldauf_2011, Horn_2015}, nor by resorting to
more sophisticated techniques, such as the use of Conformal Fermi
Coordinates (CFC) \cite{Pajer:2013ana, Dai:2015jaa, Dai:2015rda,
  Cabass:2016cgp, Cabass:2018roz}; the claimed cancellation indeed
only takes place in the unphysical {\it exactly} infinite wavelength (i.e. $k=0$) limit.
The above mentioned effects, involving long but finite-wavelength
perturbation modes, are therefore
physical and  observable in principle by future high-sensitivity experiments. In particular, primordial non-Gaussianity in single field inflation is observable even in the squeezed limit.

This paper is organised as follows. In Section \ref{long} we discuss
the transformation of the metric components under a gauge
transformations involving long-wavelength modes. In Section \ref{grad_exp} we 
show that under deformed space dilatations, the comoving curvature
$\zeta$ behaves like a 3-scalar, and no shift is present 
for any finite value of the wave-number
$k$. 
In Section \ref{Bis} the trasformation properties of
the bispectrum  under a deformed dilatation are discussed. A summary
of our main results is given in the Section \ref{conc}.
Appendix A is devoted to a discussion of the CFC approach. 
  
\section{Long perturbation modes and local rescaling}
\label{long}

Let us consider the perturbed metric around a Friedmann-Lema\^itre-Robertson-Walker (FLRW) space-time, which takes the following general form\footnote{For simplicity we neglect here the spatial
  curvature ${\cal K}$, however, the analysis can be easily extended to cover also the case $ {\cal K} \neq 0$.}
\be
ds^2 = \left(\bar g_{\mu\nu}+h_{\mu\nu} \right) dx^\mu dx^\nu =
-dt^2 + a^2 \, \delta_{ij} dx^i dx^j +h_{\mu\nu}  dx^\mu
dx^\nu \, .
\ee
The perturbation of a generic quantity ${\cal F}$ will be written as follows
\be
{\cal F}=\bar {\cal F}+{\cal F}^{(1)}+{\cal F}^{(2)}+... \, . 
\ee
Thus, at linear order, the perturbed metric
$h_{\mu\nu}$ represents a small deviation from homogeneity and isotropy. Under an infinitesimal coordinate transformation $x^\mu\to x'{}^\mu=
x^\mu+ \epsilon^\mu$ the metric components transform as
\be
\begin{split}
& \Delta h_{00}=2 \, \de_t \epsilon^0\,,\\
& \Delta h_{0i}=\partial_i \epsilon^0 -a^2 \, \de_t \epsilon^i \,, \\
& \Delta h_{ij}=-2\, a\, \dot{a}\, \delta_{ij} \, \epsilon^0
- a^2 \, \left(\partial_j \epsilon^i+\partial_i
  \epsilon^j \right) \, .
\label{gentransf}
\end{split}
\ee
As well known, cosmological perturbations can be decomposed into scalar, vector and tensor sectors, according to~\cite{Stewart:1990fm}
\be 
\begin{split}
& h_{00}^{(1)}=-2 \,\phi^{(1)} \, ,\\
& h_{0i}^{(1)}= a\, \left(\partial_i F^{(1)}+G_i^{(1)}\right) \, ,\\
& h_{ij}^{(1)}= a^2\, \left(-2 \, \psi^{(1)} \,\delta_{ij}+\partial_{ij}B^{(1)}+\partial_j C_i^{(1)}+\partial_i C_j^{(1)}+D_{ij}^{(1)}\right) \, ;
\end{split}
\label{decomp}
\ee
In particular, focusing on the scalar sector only and setting $\epsilon^i
= \de_i \epsilon$ we get
\be
\begin{split}
& \Delta \phi^{(1)}= - H \, \dot{\epsilon}^0 \, , \qquad \Delta
F^{(1)}=a^{-1} \, \epsilon^0 - a \, \dot{\epsilon} \, , \qquad \Delta
\psi^{(1)}= H \, \epsilon^0 \, , \qquad \Delta
B^{(1)} = -2 \,  \epsilon \, , \\
&\Delta {\cal F}^{(1)} = - \dot{\bar{{\cal F}}} \, \epsilon^0 \, ;
\end{split}
\label{transf}
\ee
where $\dot{f}$ denotes the time partial derivative of $f$.
It is worth to notice that the decomposition (\ref{decomp}) makes sense only if the metric perturbations have a non-trivial space
dependence;  the case of infinite-wavelength perturbation is special and will be discussed separately below. 
We will show that the standard transformation rules (\ref{transf}) are the only ones that can be
continuously deformed into the infinite-wavelength limiting case, or, in Fourier space, to the  $k \to 0$ case.

In cosmological applications, it is often useful to define the curvature of an hyper-surface defined by the condition
$S=$const., where $S$ is a four-dimensional scalar. Such a curvature ${\cal R}_S$ is defined in terms of the 3D Ricci scalar associated to the induced metric in $S$
\be
\chi_{\mu \nu} =g_{\mu \nu} + n_\mu n_\nu \, , \qquad n_\m=  \left(- g^{\alpha \beta} \de_\alpha S \de_\beta S
\right)^{-1/2}  \,  \de_\mu S , ,
\ee
as
\be
{\cal R}_{S}= R - K_{\mu \nu} K^{\mu \nu} +K^2 +2 \nabla_\nu \left(n^\mu
  \nabla_\mu n^\nu \right) -2 \nabla_\mu \left( n^\mu \nabla_\nu n^\nu
  \right) \, ;
\label{curvdef}
  \ee
where~\footnote{We denote by $\nabla$ the covariant derivative
  associated to the Levi-Civita connection.} $K_{\mu \nu}=\chi_\mu^\alpha
\nabla_\alpha n_\nu$ is the extrinsic curvature of $S$, $K=  K_{\mu
  \nu} \chi^{\mu \nu}$ its trace and $R$ is the four-dimensional Ricci scalar.
At the first order in perturbation theory we have
\be
{\cal R}_{S}=\frac{4}{a^2} \, \de^2 \left( \frac{H}{\dot{\bar{S}}} \,
  S^{(1)} - \psi \right) + \cdots \equiv {\cal R}_{S}^{(1)} + \cdots   \qquad \qquad \de^2= \delta_{ij}
\de_i \de_j \, .
\label{curvdeflin}
\ee
The geometric nature of ${\cal R}_{S}$ leads to the so-called gauge invariance of ${\cal R}_{S}^{(1)}$; namely from eq. (\ref{transf}) one
gets $\Delta  {\cal R}_{S}^{(1)}=0$. As expected, ${\cal R}_{S}^{(1)}$ is insensitive to any purely spatial transformation,
with $\epsilon^0=0$, under which neither $S^{(1)}$ nor $\psi^{(1)}$
change. We have assumed that at the background level $S$ coincides
with the
hypersurface of homogeneity of the unperturbed FLRW spacetime. Notice
that in generic coordinates, by choosing local coordinates ${y^a, \, a=,1,2,3}$ on $S$ we have that
\be
\frac{\de t}{\de y^a} = - \frac{\de_i S}{\de_t S} \frac{\de x^i}{\de
  y^a} \, .
\ee
For instance, in single-field inflation models, taking $S=const.$ as the hyper-surface whose normal is proportional to $\de_\mu\varphi$, where $\varphi$ is the inflaton field, we get
\be
{\cal R}_{S}^{(1)} \propto \de^2 \left (H \, v -\psi^{(1)} \right )  \equiv
\de^2 \zeta
\, , \qquad  v = \frac{\varphi^{(1)}}{\dot{\bar{\varphi}}} \, .
\ee
The quantity $\zeta$ is the comoving curvature perturbation and it is well known that $\zeta$ is constant in time on super-horizon scales in the case of single-field inflation. 

The case of very large-wavelength perturbations needs to be analyzed
carefully. Indeed,  consider the following transformation
\be
x^i \to x'{}^i =e^{\lambda} x^i = (1+\lambda+
\cdots)x^i \, ,
\ee
corresponding to spatial diffeomorphisms, where $\lambda$ can be space and time dependent. 
In this case, if we take $\epsilon^i= \lambda \,
x^i$ and $\epsilon^0=0$, we have
\be
\Delta h_{00} =0 \,\qquad \Delta h_{0i}= -a^2 \, x^i  \, \dot \lambda\,;
\ee
while
\be
\Delta h_{ij} = - a^2 \left(2 \, \lambda \, \delta_{ij} + x^i \, \de_j
\lambda+x^j \, \de_i \lambda \right) \,.
\label{deltah}
\ee
The 3-vector $\epsilon^i$ can be written as the gradient of a scalar
$\epsilon^i=\de_i \epsilon$,
only if
\be
\epsilon^i=\de_i \epsilon  \quad \Rightarrow x^i  \de_j \lambda-x^j \de_i \lambda =0 \, .
\label{rot}
\ee
Of course, this is the case when $\lambda= \lambda_0=const.$, where only a $\delta_{ij}$ term is generated in (\ref{deltah});
however there is a one-parameter degeneracy between the variations of $\psi^{(1)}$ and $B^{(1)}$; indeed (\ref{deltah}) is reproduced by taking
\be
\begin{split}
& \Delta_{(\alpha)} \phi^{(1)}=0 \, , \qquad \Delta_{(\alpha)} 
F^{(1)}=0 \, , \qquad \Delta_{(\alpha)}  S^{(1)} = 0 \, , \\
&\Delta_{(\alpha)}  B^{(1)} = \lambda_0(\alpha-1) x^2 \, , \qquad  \Delta_{(\alpha)} 
\psi^{(1)}=  \alpha \, \lambda_0\, .
\end{split}
\label{transconst}
\ee
Furthermore, transformation rules (\ref{transconst}) are often used by setting arbitrarily $\alpha=1$. 
This 1-parameter degeneracy can be lifted by introducing a linear dependence on the coordinates, by taking $\lambda= \lambda_0+ \lambda_1 \, x^i n_i$, with $n_i$ a constant 3-vector;  we then get
\be
\begin{split}
& \Delta \phi^{(1)}=0 \, , \qquad \Delta
F^{(1)}=0 \, , \qquad \Delta S^{(1)} = 0 \, , \\
&\Delta B^{(1)} = -\frac{\lambda}{2} x^2 \, , \qquad  \Delta
\psi^{(1)}=  \frac{\lambda}{2}\, ; \qquad \qquad \lambda=\lambda_0
+\lambda_1 \, x^i n_i
\, .
\end{split}
\label{transconst1}
\ee
As a result, a linear dependence on $\bf x$ lifts the degeneracy, by setting $\alpha=1/2$; however with this choice the spatial metric is no longer diagonal. 
The situation, in general, is even worse when $\lambda$ is a quadratic function of $x^i$; namely $\lambda=\lambda_0 + x^i n_i+
D_{ij} x^i x^j$ where ${\mathbf D}$ is a constant symmetric matrix. In this case $\Delta h_{ij}$ cannot be reproduced by any variation of $\psi$
and $B$ alone.
Such ambiguity is solved by imposing that the variation of the spatial metric $h_{ij}$ can be consistently reproduced by a change of the scalar modes in $h_{ij}$, i.e. if and only if 
$\lambda=\lambda(t,\,|\textbf{x}|)$, satisfying automatically
(\ref{rot}). 
In general, any coordinates transformation of the form $\delta x^i=
\partial_i \epsilon(t,|\textbf{x}|)$ can be written as a  {\it deformed dilatation} 
such that
\be
\lambda \, x^i= \partial_i \epsilon \;\; \Rightarrow\;\;
\lambda =\frac{\de_r \epsilon}{r} \, \qquad r =|\textbf{x}| \, .
\ee 
Thus, except for the special case $\lambda=\lambda_0=$constant, the
correct transformation rule is (\ref{transf}) and thus, being
$\epsilon^0=0$, there is no shift of $\psi$. Such a shift can be
present only in the very special case $\lambda=\lambda_0=$constant
with the choice $\alpha=1$.
In other words, there is no spatial dilatation with {\it almost constant} $\lambda(|\textbf{x}|)$ able to reduce with continuity to the $\lambda=\lambda_0$ case, simply because the large-scale limit $k \to 0$ and the smallness of the gradient term $x^i \partial_i\, \lambda$ do not commute, thereby giving rise to different transformation rules. As we will demonstrate in Section \ref{grad_exp}, this implies that terms of order $x^i \partial_i \lambda$ cannot  be neglected or confused in a gradient expansion.\\
Indeed, the choice to set $\alpha=1$ in the $\lambda_0$ case was first introduced by Weinberg \cite{Weinberg:2003sw,Weinberg:2008zzc} to prove the constancy of $\zeta$ in the large-scale limit, and later in \cite{Hinterbichler:2012nm}\cite{Hinterbichler:2013dpa} to get the single-field consistency relations. However, here the {\it gauge redundancy} approach is applied in the large-scale limit; this consists in setting the scalar variation $\Delta B$ to zero by hand, a procedure which cannot be reproduced by a standard coordinate transformation.
A typical use of such a ``would-be-shift" is the attempt to gauge away any very long-wavelength component of a perturbation in $\psi$. Consider, for instance, the case of single-field inflation and
fix the so-called {\it $\zeta$-gauge}, obtained by
setting \cite{Maldacena:2002vr}
\be
\varphi^{(1)}=B^{(1)}=0  \, ,
\label{zetag}
\ee
Thus, in this gauge, ${\cal R}_S^{(1)}=-\frac{4}{a^2} \de^2 \psi^{(1)}$ and $\zeta=-\psi^{(1)}$;  
primordial non-Gaussanity is given in terms of the 3-point function of
$\psi^{(1)}$ in this gauge. 
Suppose that we split $\psi^{(1)}$ in two parts $\psi^{(1)}= \psi_S+\psi_L$, with $\psi_S$ containing only short (high-frequency) modes and $\psi_L$ describing a
perturbation made of very long wavelengths; can we get rid of $\psi_L$ by using the shift of $\psi^{(1)}$ induced by a spatial
rescaling of coordinates, and, at the same time, remain in the $\zeta$-gauge (\ref{zetag})? 
The answer is negative, unless $\psi_L$ is genuinely constant; otherwise, instead of shifting $\psi^{(1)}$ one is going to produce a non-vanishing $B^{(1)}$, 
moving away from this gauge (\ref{zetag}). 
This is perfectly consistent with the dynamics of single-field inflation; indeed, in the gauge (\ref{zetag}), the field $\psi^{(1)}$ satisfies the equation
 \be
\ddot{\psi}-  \left(\frac{2 {\dot H}}{H}+\frac{\ddot{H}+6 H \,   \dot{H}}{2
     \plm^2 \, \dot{H}^2 }+3 H\right) \dot{\psi} + \frac{1}{a^2} \, \de^2
 \psi=0 \, .
\label{psieq}
 \ee
In such a gauge, therefore, a rescaling of the coordinates should send a solution
of (\ref{psieq}) into a new solution. Namely, considering an arbitrary shift $\Delta \psi^{(1)}= \lambda(\bf{x})$,
$\lambda$ should solve (\ref{psieq}). Actually this is the case if $\de^2 \lambda=0$; thus $\lambda= \lambda_0 +  \lambda_1 \, x^in_i$, thereby confirming the above conclusion. 
In the Fourier basis, this means that $\lambda_k$ is a delta-function
with support in $k=0$  that this is precisely what is needed in
the proof of the generalised consistency relations \cite{Hinterbichler:2013dpa,Hui:2018cag}.

\section{Deformed dilatations and gradient expansion}
\label{grad_exp}
In this Section we show that, whatever the selected range of scales is, a deformed dilatation can be described via standard gauge transformation rules, avoiding the $\psi$ shift in the large-scale limit.
Consider a deformed dilatation of the form
\be
x^i \to {x'}^i= x^i + \lambda(x) \, x^i \, , \qquad \qquad \lambda=
\frac{1}{(2 \, \pi)^{3/2}} \int d^3 k \, e^{i
  \textbf{k}\textbf{x}} \lambda_k \,,
\label{ddil}
\ee
where usually $\lambda_k$ is taken to be the filtered Fourier transform of some comoving curvature mode $\zeta_k$. Often the window function $W$ is supposed to select  the ``long mode" part of $\zeta$, namely 
\be
\lambda_k= W_k\, \zeta_k\, ,
\ee
that should be gauged away.
A widely used choice ~\footnote{See for instance
  \cite{Pajer:2013ana}, were a transformation to Conformal-Fermi-Coordinates is approximated by a deformed spatial dilatation with $W$ taken as an Heaviside step function.} for $W_k$ is the Heaviside step function centred on a particular scale $k_c$, 
\be
\label{window}
W_k=\theta\left[\frac{1}{H}(k_c-k)\right]\,, 
\ee
where  $k_c/H \ll 1$ at the time of interest.
Naively, one could argue that performing such a ``dilatation'', the off-diagonal terms of the transformed spatial metric $\sim x^j \partial_{i} \lambda$ in (\ref{transconst}) are
negligible, at leading order in a gradient expansion. 
However, integrating by parts we get
\be
 x^i \partial_j \lambda=-\frac{1}{(2\, \pi)^{3/2}} \, \int
 d^3 k \, e^{i\, \textbf{k} \cdot \textbf{x}} \partial_{k^i} \left(k^j \,\lambda_k \right)+ \text{BT} \, .
\label{der}
\ee
The window function is such that the boundary term BT can be set to zero, given that $W$ selects long modes only. We have taken both $\lambda_k$ and $W_k$ functions of $k=|\bf{k}|$; as a result  (\ref{rot}) is satisfied. Here differentiation with respect to $k$ will be denoted by a prime.
Thus, in Fourier space
\be
\begin{split}
\left(\Delta {h_k}\right)_{ij} &= - a^2\left[2 \, \lambda_k \,
  \delta_{ij}- \partial_{k^j} \left(k^i\, \lambda_k\right)-
  \partial_{k^i} \left(k^j\, \lambda_k\right)\right]\\
&=  2 \, a^2 \, \frac{k^i k^j}{k} \left(\lambda_k\right)^\prime \,.
\end{split}
\ee
Unless $W_k \sim \delta(k)$, there is no term proportional to $\delta_{ij}$, while $\left(\Delta {h_k}\right)_{ij}$ is reproduced by
\be
\Delta B_k =-\frac{\Delta h_{ij}}{a^2\, k^i\,k^j}= - \frac{2}{k}
\lambda_k^\prime \, ;
\label{trB}
\ee
where $\lambda_k' = \de_k \lambda_k$.
No shift of $\psi$ is present and, once again, we recover the general transformation rule (\ref{transf}). Indeed, by using (\ref{der}) we obtain $\epsilon$ such that $\de_i \epsilon = \epsilon^i$ 
is given in Fourier space by
\be
\epsilon_k = \frac{\lambda_k^\prime}{k} \, ;
\ee
thus (\ref{trB}) is exactly the Fourier transform of (\ref{transf}), as expected. Despite the filtering procedure which drops short modes,
the transformation rule (\ref{transf}) is not altered.
Our result is in contrast with the transformation rule of the curvature perturbation and definition advocated
in~\cite{Cabass:2016cgp}, where, instead of the comoving curvature perturbation, a quantity more similar to the local number of
 e-fold is defined\footnote{Actually, the local number of e-folds contains terms of order $\partial^2 F$, coming from a proper time integration along the fluid world-line, with relative terms lying on the initial hyper-surface \cite{Matarrese:2018qqo}.} 
\be
\zeta^{(\text{alt})}=\delta N =\frac{1}{6} \, \frac{ h_{ii}}{a^2}=-\psi +\frac{1}{6}
\, \partial^2 \, B \, .
\label{CFC}
\ee
While the quantity  $\delta N$, which coincides with $\zeta$ in the comoving gauge\cite{Sasaki:1995aw, Wands:2000dp, Lyth:2004gb, Abolhasani:2018gyz},
is often assumed to be still approximately equal to $\zeta$ in
CFC and other gauges  where the term $\partial^2 B$ is assumed to be negligible in the gradient expansion. 
The $\partial^2 B$ in CFC is very peculiar, indeed it compensates the diagonal metric term $\psi\, \delta_{ij}$ leading to $g_{ij}\sim x^2  \, \de^2 \zeta$. However, we have just shown that such gradient terms cannot be neglected and in Fourier space it does not matter how close to zero the $x^i$ value is. Let us also stress that  $\zeta^{(\text{alt})}$ is rather different from the comoving curvature perturbation $\zeta$; most crucially it is not gauge invariant! 
Indeed, in Fourier space, we have that, under the transformation (\ref{ddil}),
\be
\label{one} 
 \zeta^{(\text{alt})}_k \to \zeta^{(\text{alt})}_k +\frac{1}{3}
\,k \, \lambda_k' \,
\ee
where again there is no shift term, and in addition 
and the gauge invariance is simply lost. 
In Appendix (\ref{CFC_appendix}) it is shown that the coordinate transformation that connects CFC to comoving coordinates can be
written as a deformed dilatation with a suitable $\lambda$.
The shift of $\psi$ is crucial in arguing that the long mode can be gauged away by transforming the primordial bispectrum, or any
N-point function obtained by a {\it local measurement}
\cite{Dai:2015rda}, from comoving to CFC coordinates and then canceling the
leading term of the Maldacena consistency relation \cite{Pajer:2013ana,Cabass:2016cgp,Bravo:2017gct}. Contrary to what we have found, it is often
claimed that  under a spatial  ``long'' deformed dilatation transformation where
$\zeta^{(\text{alt})}$~\footnote{We stress once again that such a quantity
is not the gauge-invariant $\zeta$ unambiguously defined in any coordinates
as being proportional to the Ricci scalar of the hyper-surface
orthogonal to the inflaton velocity in single field inflation.} is
split into a long and a short part, while the ``long''
$\zeta^{(\text{alt})}_L$ shifts by $\lambda_L$, the short part $\zeta^{(\text{alt})}_S$ is a
genuine scalar quantity. In our analysis we do not find such a transformation property. 

The first application of a constant dilatation was given in~\cite{Weinberg:2003sw}\cite{Weinberg:2008zzc} as a tool to show the constancy of $\zeta$ under mild assumptions. A constant dilatation represents the residual gauge ambiguity in the Newtonian gauge of an unperturbed FLRW solution. Such a pure-gauge mode with $k= 0$\footnote{The crucial difference from standard transformations is to impose a gauge redundancy in the $k\equiv 0$ case} can be promoted to a physical (adiabatic mode) perturbation by enforcing that it solves the subset of Einstein's equations that are trivial at $k= 0$ (for instance in the Newtonian gauge this set corresponds to the $ij$ Einstein's equations.).
In this sense a non-physical gauge mode that corresponds to a constant spatial dilatation is promoted to the physical adiabatic mode 
\be
\zeta \mid_{k \to 0}=-\lambda_0 - \frac{H}{a}\, C \, ;
\ee  
where, typically, the constant $C$ corresponds to a decaying mode.
Similarly, the Maldacena consistency relation and their extensions can be derived~\cite{Creminelli:2011rh,Hinterbichler:2012nm,Hinterbichler:2013dpa,Hui:2018cag}
by promoting the redundancy of the $\zeta-$gauge to a full-fledged adiabatic mode, extending the transformation including $3$-special
conformal transformations in addition to constant dilatations.\\
The Maldacena consistency relation relates the 3-point function of the
comoving curvature perturbation to the 2-point function of the same
quantity. It is important that such correlation functions are scalars
under a change of the coordinates of the hyper-surface, see
(\ref{curvdef}) and (\ref{curvdeflin}), namely
\be
{\cal B}(\vec{x}_1, \vec{x}_2, \vec{x}_3) =\langle \zeta(\vec{x}_1) \,
\zeta(\vec{x}_3) \,  \zeta(\vec{x}_3) \rangle \to {\cal B}(\vec{\tilde{x}}_1,
\vec{\tilde{x}}_2, \vec{\tilde{x}}_3) ={\cal B}(\vec{x}_1, \vec{x}_2, \vec{x}_3)  \, ;
\label{bitransf}
\ee
a similar relation holds true for the 2-point function\footnote{We are
  interested on super-horizon scales where, at least when the Weinberg
  theorem applies, the time dependence in ${\cal B}$ is negligible (for simplicity of notation, only spatial coordinates will be shown).}.
This relation is a consequence of the geometric nature of $\zeta$ and of the general covariance of the action
  describing gravity and the inflationary sector. According to \cite{Maldacena:2002vr}, in the squeezed limit, where one of the three momenta is much smaller than the others, the primordial bispectrum assumes the following  form
\be
\label{con_rel}
 \langle \zeta(k_1)\zeta(k_2)\zeta(k_3) \rangle\mid_{{}_{k_3\to
     0}}=-(2 \, \pi)^3\,
 \delta(\textbf{k}_1+\textbf{k}_2+\textbf{k}_3)\,(n_s-1)\, P(k_3)
 \,P(k_1)\, .
 \ee
 If one accepts the transformation property (\ref{bitransf}), it
 seems of little physical interest the choice of coordinates used in
 the computation of ${\cal B}$. Indeed, once ${\cal B}$ is found to be non-vanishing
in a set of coordinates, the same will be true in any other
set of coordinates. In the non-linear case, taking the expectation value on the Bunch-Davis
vacuum state does not commute with the  transformation of
$\zeta$.

By using CFC,  in \cite{Pajer:2013ana,Cabass:2016cgp} it was
argued that actually {\it only} the short-wavelength part $\zeta_s$ of
$\zeta$ can be defined as a scalar and, at the
leading order, one has
\be
\tilde \zeta_s(x)= \zeta_s(x)-\lambda \,x^i\,\partial_i\,\zeta_s(x)\,;
\label{czeta}
\ee  
where $\tilde \zeta_s$ is the short part of $\zeta$ in CFC and
$\lambda$ is the deformed dilatation which relates CFC and comoving
coordinates, see \ref{CFC_appendix} and \cite{Pajer:2013ana,Dai:2015rda}.
Their claim can be summarised by the following relation valid {\it in the
squeezed limit}
\be
\label{P_res}
\Delta {\cal B}= \tilde {\cal B}(x_1,\, x_2,\, x_3)-{\cal B}(x_1,\, x_2,\,x_3 ) \equiv -{\cal B}(x_1,\, x_2,\, x_3)\,;
\ee
thus $\tilde {\cal B}_{\text{squeezed}}=0$.
According to our result, actually 
\be
\Delta {\cal B} \equiv 0 \, ,
\label{bitransf1}
\ee
in agreement with (\ref{transf}). Two independent arguments in favour of (\ref{bitransf1}) will be given in Section \ref{Bis}.
By using the {\it in-in formalism}, one can show that the effect of the transformation to CFC changes the Lagrangian by a total
derivative. Alternatively, by direct computation of the leading order term in perturbation theory, $\Delta {\cal B}$ in Fourier space is found to be just a vanishing boundary term.

In summary, we have argued that single-field primordial
non-Gaussianity is observable even in the squeezed limit, and thus the
Maldacena consistency relation is physical and observable in
principle. Such a result is in agreement with the results of \cite{Domenech:2016zxn,Abolhasani:2018gyz},
according to which the local $f_{\rm NL}$ is invariant under a non-linear
field redefinition, and frame-independent \cite{Gomez:2020gqa}.  A similar result was achieved in \cite{Giddings_2011} where it is argued that semi-classical relations are physical away from the exact $k_L =0$ limit, including the consistency one, as it induces statistical anisotropy in the powerspectrum of short modes. Note that, the Maldacena consistency relation can be
  derived by using only the residual dilatation symmetry in the
  $\zeta$-gauge. The sole effect of  a change of time  or a more complicated
  special conformal transformation is  to give rise  to higher order
  correction in the slow roll-parameters or a gradient
  correction in the consistency relation
  \cite{Hinterbichler:2013dpa}\cite{Creminelli:2013mca}\footnote{In
  \cite{Creminelli:2013mca}, the change of time is important to find
  the gauge redundancy in the Newtonian gauge.} .

We conclude this section by taking into account another example where space-dependent dilatations were widely used in the literature
\cite{dePutter:2015vga,Cabass:2018roz} to study the scale-dependence of the dark matter halo bias, in connection with possible signatures of primordial non-Gaussianity.
In particular, disagreement persists in the literature about the possible total or partial cancellation of the effective
 $f_{\rm NL}^{\rm GR}=-5/3$ contribution. Looking at works where CFC or simpler dilatation transformations were used, the following two ingredients are always present 
\begin{enumerate}
\item the off-diagonal terms of the metric are neglected, invoking a gradient expansion; however, as we have just shown, in Fourier space they are of the same order as $\zeta$ itself and fundamental for the covariance of the theory, hence they cannot be neglected;
\item an extensive use of the separate universe approach is always present. According to this approach the metric is diagonal on very large scales, 
ensuring equivalence (in particular for the halo-bias) between comoving and synchronous gauges and between the $\delta N$ local number of e-folds and the comoving curvature $\zeta$. Finally, the comoving curvature is related to the energy-density field.
\end{enumerate}
It seems that the presence of these two ingredients is sufficient to
claim the cancellation of any single-clock primordial 
$f_{\rm NL}$ contribution in local measurements \cite{Dai:2015rda,Cabass:2018roz}. 
However, in the light of our  results some doubts are in order. 
The definition of $\zeta$ in CFC coordinates as the  trace of the spatial metric $g_{ij}$ and relating it to the energy-density can
be dangerous. Such a quantity, which is of second order in the CFC expansion, is not a genuine 3-scalar invariant; see eq. (\ref{one}).
On the other hand, the energy density is well known to be related to the gravitational  potential $\psi$ via the Poisson equation, which in a generic gauge reads 
\be
\label{rho}
8\, \pi\, G \,\rho=\frac{2}{a^2} \, \nabla^2 \psi-\frac{6}{a^2}\, H\, \dot{\psi}+\frac{H}{a^3}\, \left(a\,\nabla^2 \dot{B}-2\, H\, \nabla^2 F\right)\,.
\ee
eq. (\ref{rho}) is perfectly invariant under the deformed dilatation because the $B$ gauge change is compensated by the scalar $F$ transformation. This is sufficient to conclude that the  
long mode with finite momentum $k$ cannot be gauged away by a spatial
dilatation, exactly as shown for the consistency relation.
Sometimes the cancellation is motivated by resorting to the Equivalence
Principle. Now $\zeta$ plays the role of a gravitational potential and
indeed, according to our result it can be shifted by a constant, as a
consequence of the ambiguity present at $k=0$. Such an ambiguity is
present in the case of $\lambda(x) = \lambda_0 + \vec{x} \cdot \vec{b}$, see
(\ref{ddil}) and the end of the previous section. However, when $\vec{b} \neq 0$, (\ref{rot}) is not
satisfied and thus such a $\lambda$ is not a coordinates
transformation that affects the scalar sector. As soon as $\lambda$ is
at least a quadratic function of $\vec{x}$, tidal effects kick in and
no shift of $\zeta$ is present. In this sense, our result is in
full agreement with the Equivalence Principle.

\section{Bispectrum gauge transformation under deformed dilatations}
\label{Bis}

In this Section, we show that under a deformed dilatation, the bispectrum of a generic three-scalar is unchanged, in the sense that 
$\Delta {\cal B}=\tilde {\cal B} (x_1,\,x_2,\,x_3)-{\cal B}(x_1,\,x_2,\,x_3)\equiv 0 $. 
Tree-level correlation functions can be computed in two ways: by using the in-in formalism in the interacting picture or equivalently by
solving the classical equation of motions at the required order in perturbation theory with free-field initial conditions, see for instance \cite{Weinberg:2005vy}.
In the in-in formalism, correlations of fields are computed by relating the fields in the interaction picture (free fields) with Heisenberg
picture fields perturbatively; thus non-linearities are encoded in such a relation. 
In the second approach fields are decomposed with creation and annihilation operators with non-linear modes. \\
Let us start with in-in formalism. It is sufficient to analyse the
extra terms in the action, arising from the non-linear coordinate transformation (deformed dilatation) which 
connects comoving gauge, where $v$ and $B$ are set to zero, with CFC-like reference frame
\be
\label{Nonlinear_dil}
\tilde x^i= e^{\lambda}\, x^i\,, \qquad g_{ij}=a^2 \, e^{2 \, \zeta}\,\delta_{ij}\,.
\ee
The action is invariant and can be written in ADM form as \cite{Maldacena:2002vr}
\be
\label{Action}
S=\int \, d^4x \, \sqrt{h} \, N\,\left[ R^{(3)}+K_{ij} \,
  K^{ij}-K^2+{\cal L}_m\right] \equiv \int \, d^4x \, \sqrt{h(x)} \,
{\cal S}(x)\,.
\ee
where $h$ is the spatial metric determinant, $N$ is the shift and $K_{ij}$ is the extrinsic curvature tensor of $t=constant$, while
${\cal L}_m$ is the Lagrangian for the inflaton $\phi$. In the comoving gauge, the $t=constant$ hyper-surface coincides with
$\phi=$const. hyper-surface.
The 3-scalar in (\ref{Action}) can be written as 
${\cal S}$
\be
\label{tr_prop}
\tilde {\cal S}(\tilde x) \equiv {\cal S}(x)= \bar {\cal S}(t)+{\cal S}^{(1)}(x)+{\cal S}^{(2)}(x)+\dots\,. 
\ee
Defining the quantity
\be 
\Delta_{S}=\sqrt{\tilde h (x)}\, \tilde {\cal S}(x)-\sqrt{h(x)}\,\, {\cal S}(x)\,,
\ee
the bispectrum variation in-in formalism is given by
\be
\begin{split}
\label{DeltaB}
\Delta {\cal B}&= \langle \tilde \zeta (t\, , x)^3 \rangle-\langle \zeta (t\,
,x)^3 \rangle \\
&=i\,\int_{t_0}^t \, dt' \, \langle \left[\zeta^3(t,\,x),\,\int
\,d^3 x \, \Delta_{S}^{(3)}\right] \rangle\,,
\end{split}
\ee
the second equality can be obtained by using the first-order gauge invariance \footnote{As we have previously
  shown, excluding the very special case of a constant $\lambda$, this is the case.} of $\zeta$:
$\tilde{\zeta}^{(1)}(x)=\zeta^{(1)}(x)$  and  
\be 
L_{int}=-H_{int}=\int \,d^3 x \, \sqrt{h}\, {\cal S}.   
\ee
Now, under a generic redefinition of spatial coordinates $\delta x^i= \tilde{x}^i-x^i$, we get the following transformation
\be
\begin{cases}
& \tilde {\cal S} (x) = {\cal S}(x)-\delta x^i \, \partial_i \left( {\cal S}^{(1)}(x)+{\cal S}^{(2)}(x) \right)-\frac{1}{2}\, \delta x^i\, \delta x^j \, \partial_{ij} {\cal S}^{(1)}(x)+\delta x^j \, \partial_j \left(\delta x^i \, \partial_i {\cal S}^{(1)}\right)\,;\\
& \sqrt{\tilde h}(x)= a^3 +a^3 \, \left[3 \, \zeta- \partial_i \left(x^i\, \lambda\right)\right]+\frac{1}{2} \, a^3 \, \left[9 \, \zeta^2- 6 \, \zeta\, \partial_j \left(x^j \, \lambda\right)-6\, \lambda \, x^j \,\partial_j \zeta\right]\\
&\qquad\qquad +\frac{1}{2}\, a^3\, \left[9\, \zeta^3 -9\, \zeta^2\, \partial_j \left(x^j\, \lambda\right) -18\, \zeta\, \lambda\, x^j\, \partial_j\zeta\right]+O(\lambda^2\,, \lambda^3)\,.
\end{cases}
\ee 
For simplicity, we have omitted terms quadratic and cubic in $\lambda$,
coming from the transformation of the determinant of the spatial
metric $h$. All the $\lambda^2$ terms in the quadratic action can be canceled, by integrating by parts, and the same reasoning applies to
$\lambda^2-\lambda^3$ vertices in the cubic action
\footnote{Note that being $\lambda$ defined by long modes only,
  $\lambda^2-\lambda^3$ vertices should imply triangles with two and
  three squeezed momenta that are not relevant in the squeezed limit.}. 
Thus, the change of the spatial coordinates induces the following variation $\Delta_{S}$ up to the third order
\be
\label{dletaS_res_back_1_2}
\begin{cases}
&\bar \Delta_S=0\,,\\
&\Delta_{S}^{(1)}=
a^3 \,\bar S(t)\, \partial_i \left(\lambda\, x^i \right)\,,\\
&\Delta_{S}^{(2)}= 
 -\frac{1}{2}\, a^3 \, \partial_i\,\left[2\,\lambda \, {\cal S}^{(1)}\, x^i +6\,\bar {\cal S}(t) \,\lambda\,\zeta\, x^i\right]\,,\\
& \Delta_S^{(3)}=-\frac{9}{2}\, a^3 \,\bar {\cal S}(t)\,\partial_i \left(\zeta^2 \, \lambda\, x^i\right)-3 \, a^3 \, \partial_i \left(\zeta\, \lambda\, {\cal S}^{(1)}\right)-a^3 \, \partial_i \left(\lambda\, \bar{\cal S}^{(2)}\, x^i\right)\,,
\end{cases}
\ee
we have omitted all the quadratic and cubic terms in $\lambda$. 
The final result is that, for all the relevant vertices, we get that the variation of the cubic Lagrangian is just a boundary term and then, 
substituting in eq. (\ref{DeltaB}), we get
\be
\Delta {\cal B} \equiv 0\,.
\label{var}
\ee
The above analysis can be also extended to the case
  where a change of time is considered; the result will be again (\ref{var}).\\
Let us now reproduce the same result in a different way. 
The first step consists in considering that the $\zeta$ curvature is a 3-scalar, i.e.
\be
 \tilde{\zeta}(\tilde x)= \zeta(x) \;\; \Rightarrow \;\;\tilde \zeta
 (x)=\zeta (x)-\lambda \,x^i \,\partial_i\, \zeta (x) + \cdots \,. 
\label{corr}
 \ee
The variation of the spectrum due to the above non-linear transformation can be computed at the leading order as a correction
proportional to a 4-point function, which is already non-vanishing for
free fields. From (\ref{corr}), we get in Fourier space
\be
\label{nonlin_deltaB}
\Delta {\cal B} =  -\langle \zeta_{k_1}\zeta_{k_2} \left(\lambda\, x \cdot \partial\, \zeta\right)|_{k_3}\rangle - \langle \zeta_{k_1} \left(\lambda\, x \cdot \partial\, \zeta\right)|_{k_2} \zeta_{k_3}\rangle -\langle \left(\lambda\, x \cdot \partial\, \zeta\right)|_{k_1} \zeta_{k_2}  \zeta_{k_3}\rangle \equiv \sum_{i=1}^3 \, {\cal B}_i \, ;
\ee
where by definition
\be
\begin{split}
&\left(\lambda\, x \cdot \partial\, \zeta\right)|_{k} =\int \frac{d^3 x}{(2\,\pi)^9}\,e^{-i\, \textbf{k}\,\textbf{x}}\left( \Pi_{i=1}^2 \, \int d^3 p_i\, e^{i \, \textbf{p}_1 \, \textbf{x}}\, x^i\partial_i \left[ e^{i \, \textbf{p}_2 \, \textbf{x}}\right]\,\zeta_{p_2} \, \lambda_{p_1}\right)\,, \\
&\zeta_k = v_k \, a_{\textbf{k}}+v_k^* \, a_{\textbf{k}}^{\dagger}\,,\qquad \lambda_k =v_k \,W_k \, a_{\textbf{k}}+\bar v_k^* \,W_k\, a_{\textbf{k}}^{\dagger}\,.
\end{split}
\ee
Focusing on the first term in relation (\ref{nonlin_deltaB}); we get, at the leading order in perturbation theory,
\be
 \begin{split}
\Delta {\cal B}_3  &=\int \frac{dx^3}{(2\, \pi)^9}\, e^{-i \, \textbf{k}_3 \,\textbf{x}}\, \Pi_{i=1}^2\, d^3 p_i\, e^{i \, \textbf{p}_1 \, \textbf{x}}\, x^i \, \partial_i \left[e^{i \, \textbf{p}_2\, \textbf{x}}\right]\,\langle\zeta_{k_1}\,\zeta_{k_2}\, \lambda_{p_1}\, \zeta_{p_2}\rangle \\
 &= -\int \frac{d^3 p_2}{(2\, \pi)^6}\,\partial_{p_2^j}\left[p_2^j\, \langle \zeta_{k_1}\,\zeta_{k_2}\, \lambda_{(k_3-p_2)}\, \zeta_{p_2}\rangle \right]\,,
\end{split}
\ee
where the second line was obtained integrating by parts.
Integrating ${\Delta {\cal B}}_3$ and applying the Wick's theorem, we get a vanishing boundary term considering both the presence of the window function $W_k$ and the Dirac delta term which centers $p_2$ on one of the finite momentum $k_i$. 
The same procedure applies for the other $\Delta {\cal B}_i$ terms, obtaining that the total variation $\Delta {\cal B}$ vanishes as expected. 

We conclude this Section by stressing how crucial the first-order transformation properties of the field whose bispectrum is analyzed, are. The only way to cancel the $f_{\rm NL}$ term in the squeezed limit is to allow the long modes to shift. Indeed, consider a general field $\chi$ such that at linear order
\be 
x^i \to \tilde x^i=(1+\lambda)\,x^i \;\;\Rightarrow\;\; \tilde\chi_k= \chi_{k} \,\left(1-W_k\right)\,. 
\ee
where $W$ is the above mentioned window function which isolates long modes, i.e. simply
\be 
\tilde \chi_{k_{L}}\equiv 0\,, \qquad \tilde \chi_{k_{S}}= \chi_{k_{S}}\,,
\ee
Being the action corrected thanks to total derivative terms, we trivially get
\be
\langle \tilde \chi_{k_{L}} \, \tilde \chi_{k_{S\,2}}\,\tilde \chi_{k_{S\,3}}\rangle \sim \langle \left[\tilde \chi_{k_{L}} \,  \chi_{k_{S\,2}}\,\chi_{k_{S\,3}}\,,\; \sqrt{h}\, {\cal S}\mid_{{}_{k\,k'\,k''}}\right] \rangle\,.
\ee 
The previous relation can be zero if and only if $\chi$ shifts, i.e. $\chi_{k_{L}}\equiv 0$, and in this sense the importance we gave to this point in the previous Section can be understood. 
Note that here the $\chi$ field can be replaced by $\zeta$, $\psi$ or the energy-density $\rho$, or any other field to which in the literature such a property of transformation was applied. In this sense, the $f_{\rm NL}$ term can be removed only in the {\it strictly} $k_L\equiv 0$ {\it limit}, where a redundant gauge transformation can be implemented, however this limit has {\it no physical meaning} in 
cosmological observables.\\
Furthermore, if one chooses as $\chi$ field the local number of e-folds $\delta N$, instead of $\zeta$, as done in CFC coordinates, using eq. (\ref{one}) one should get that the result is window-function-dependent near the reference scale $k_L \sim k_c$ (usually, patch-dependent), it is even singular if $W_k$ is taken to be a Heaviside step function as in eq. (\ref{window}). This makes the claimed $f_{\rm NL}$ cancellation, even less robust in these coordinates.  
       
\section{Conclusions}
\label{conc}

The detection of primordial non-Gaussianity is one of the most important
avenues of modern cosmology and forthcoming probes will be able to
significantly improve our knowledge of inflation in the Early Universe. On one hand, in the squeezed limit, the amount of non-Gaussianity
is completely fixed in a model-independent way in the case of standard single-field inflation; on the
other hand, concerns about the physical observability of such a limit have
been advocated. In the debate, it is crucial to determine how a very long
perturbation affects the quantities of physical interest.
We have reanalysed the transformation properties of cosmological
observables and, in particular, of the curvature perturbation $\zeta$
and its related correlation functions. 
Our results imply that, excluding the case of infinitely long-wavelength (hence non-physical) perturbations, $\zeta$ is a genuine geometrical quantity and in particular
it is gauge invariant at first order in
perturbation theory. By using a deformed dilatation of spatial
coordinates, no shift in $\zeta$ is found,  no matter what window function is used
to filter out short modes. A gauge ambiguity exists only in the strictly
$k \to 0$ limit.\\
Let us recap briefly the argument here.
The key property used in the cancellation is that, under a coordinate transformation, the long-wavelength part of
$\psi(x)=\psi_S(x)+\psi_L(x)$ shifts, or equivalently $\psi_L$ is absorbed as a local contribution to the scale-factor. 
From our analysis in section \ref{long}, such a shift does not exist
for a physical perturbation. There is an ambiguity when one
considers an infinitesimal spatial coordinate transformation of the kind $\delta
x^i = \lambda\, x^i$, used to gauge away the long-wavelength part $\psi_L$.\\
Imagine now to expand (in a gradient expansion) $\lambda$ in powers of $x^i$,
$\lambda= \lambda_0 + n_i x^i + D_{ij} x^i x^j + \cdots$.
We have shown that at zeroth order, there is a 1-parameter ambiguity in
the transformation properties of the scalar part of the metric,
see (\ref{transconst}). One can choose the parameter such that $\psi$
shifts, although this is not the only possibility, and then in the new
coordinate system $\tilde \psi_L=0$. When higher-order
terms in the expansion are considered, once a scalar-tensor decomposition is set, the ambiguity disappears and the
the only possibility is to leave $\psi$ unchanged, while the off-diagonal part of
$g_{ij}$ is modified, recovering the standard transformation property
(\ref{transf}).  As a result, only unphysical, genuinely x-independent, constant modes can be gauged away. More explicitly, this means that a perfectly
constant dilatation can be used to gauge away a truly constant
perturbation $\psi$; in Fourier space this is equivalent to take the $\lambda_k$
proportional to a delta function with support in $k=0$. Consider now the extension of a  constant dilation to long physical modes
by introducing a suitable window function, as implicitly done in the
literature. Take for instance a Heaviside step window function $W_k= \theta\left[H^{-1}(k_c-k)\right]$, as done in~\cite{Pajer:2013ana};
where $k_c$ will be an unspecified cutoff scale, on the edge of the
short-long region.  If one performs the following coordinate transformation
\be
\tilde{x}^i = x^i+ \zeta_L(x)\, x^i\,, \qquad \zeta_L(x)=\frac{1}{(2\,\pi)^{\frac{3}{2}}} \, \int dk^3 \, e^{i\, \textbf{k}\textbf{x}} \, W_k \, \zeta_k\,,
\ee
in order to gauge away the long part of $\zeta$ defined as
$\zeta=-\psi+H\, v$. As commonly stated, one should obtain the transformation
\be
\label{long_shift}
\zeta_L \to \tilde \zeta_L=0\,, 
\ee
at {\it zero} order in a gradient expansion.
However, when one considers the $\partial_i \zeta_L$ terms, the correct gauge transformations are 
\be
\label{correct}
\tilde \psi= \psi ,  \qquad \qquad
\tilde B = B-\frac{2}{k} \, \partial_k (W_k\, \zeta_k)\,.
\ee 
Indeed, the metric tensor term $\partial_i \partial_j B$ is of order
$\zeta_{k}$. Thus, once again, working with a physical mode, no shift
is found. The crucial point necessary to get (\ref{long_shift}) is
imposing that  $\tilde B-B=0$.
The subtlety here is that, for more standard coordinates transformations,
one can always neglect $\partial_i \de_j  B$ on large scales (doing
so $\partial_i \partial_j (\tilde B-B) \to 0$), but this is simply not the case here. 
Thus, $\tilde{\psi}-\psi=0$
and, given that at the relevant order in perturbation theory the
velocity potential does not change going from comoving coordinates to
CFC, namely $\tilde{v}-v=0$, we get that the curvature perturbation cannot change
\be
\tilde \zeta-\zeta =0\,,
\label{OK}
\ee
contrary to (\ref{long_shift}).
As one can see from eq. (\ref{transconst}) , one can always
choose $\alpha=0$ and fixing the ambiguity present at $k=0$ such that
$B$ actually transforms, making the gradient expansion continuous.
Thus, we believe that any attempt to separate ``local'' from ``global''
effects should not introduce a discontinuity in the transformation
properties of the metric.
The bottom line is that no shift exists for $\psi$. Thus, a long (but
not infinitely long!) physical mode {\it cannot be gauged away, even
  locally}. It is worth to notice the choice of what is called $\zeta$
in CFC coordinates, $\zeta^{(\text{alt})}$ in (\ref{CFC}), is not
gauge invariant, see (\ref{one}),
and does not coincide with $\zeta$ at large scale. Of course, the
same discussion can be extended to any quantity directly related to
$\zeta$, such as the energy density $\rho_k \sim k^2 \zeta_k$.
Last but not least, the transformation property  (\ref{CFC})  of
$\zeta_{CFC}$ depends strongly on the window
function $W_k$ used;  such a term
of order $\partial_k W_k$ would then affect any n-point correlator, in
the region close to $k= k_c$.\\
The impact on the bispectrum of $\zeta$ is relevant: we have found that, even in the squeezed limit, no
cancellation of primordial non-Gaussianity takes place. Thus, the Maldacena consistency relation still represents an important feature
 to tame the zoo of inflationary models, according to the pattern of symmetry breaking during inflation. 
Interestingly, similar techniques, based on deformed spatial dilatations, where used to study the general relativistic scale-dependent contribution to the bias of dark matter halos, claiming
the cancellation of the so-called $f_{\rm NL}^{\rm GR}=-5/3$ term in local measurements. Our analysis implies that also such a term is unaffected by deformed spatial dilatations.\footnote{We will discuss this issue in more detail in a future publication.}
Hence, these effects are physical and observable in principle by future high-sensitivity experiments.

\section*{Acknowledgements}
We thank Nicola Bartolo, Daniele Bertacca, Marco Bruni, Guillem Domenech, Martin Sloth, Raul Jimenez, Lam Hui, Juan Maldacena, Jorge Nore\~{n}a, Alvise Raccanelli, Antonio Riotto, Misao Sasaki and Licia Verde for useful comments and discussions. SM acknowledges partial financial support by ASI Grant No. 2016-24-H.0.

\appendix

\section{CFC vs. space-dependent dilatation}
\label{CFC_appendix}

The coordinates transformation that relates CFC and comoving
coordinates is a special case of deformed dilatation of Section
\ref{grad_exp}. The perturbed conformal metric $\tilde h_{\mu\nu}$
is related to $h_{\mu\nu}$ by 
\be
\tilde h_{\mu\nu}=\frac{1}{a^2}\,h_{\mu\nu}\,. 
\ee
Conformal Fermi Coordinates can be  described by the following tetrads
in comoving coordinates \footnote{For details see \cite{Dai:2015rda}}:
\be
\begin{split}
& e_0^{\mu}= \frac{1}{a} \left(1+\frac{1}{2} \tilde h_{00}, V^i\right)\\
&  e_i^{\mu}= \frac{1}{a} \left(0,\delta^{i}_j-\frac{1}{2}\tilde h^i_j\right)
\end{split}
\ee 
Neglecting second-order perturbations, the cubic transformation from the comoving to the CFC frame has the following form:
\be
\begin{split}
\Delta x_F^k  =&\Delta x^k-\bar \Delta x^k\, \psi |_p-\frac{1}{2}\,\bar \Delta x^i\bar \Delta x^j\, \left(\delta^i_k \,\partial_j \psi+\delta^j_k\, \partial_i \psi-\delta^i_j \, \partial_k \psi \right)|_p\\
&-\frac{1}{6} \,\bar \Delta x^i\bar \Delta x^j\,\bar \Delta x^l \left(\delta^i_k \,\partial_{jl} \psi+\delta^j_k\, \partial_{il} \psi-\delta^i_j \, \partial_{ki} \psi \right)|_p
\end{split}
\label{expC}
\ee
where $\Delta x^\mu= x^{\mu}(\tau)-p^{\mu}(\tau)$, is the deviation from a central worldline $p^{\mu}$, and $\bar\Delta x^\mu$ its background value.\\
We can take $\psi(x) = \psi(|\bar\Delta x|)$ and $\partial_j \tilde
\psi|_p= \partial_j \psi |_{\bar\Delta x=0}$ and, thus, without loss of generality,
\be
\begin{split}
\label{CFC_form}
\Delta x_F^k \equiv x_F^k= &x^k-x^k{\cal V}(x)- x^k\, \psi |_0-\frac{1}{2}\, x^i\,x^j\, \left(\delta^i_k \,\partial_j \psi+\delta^j_k\, \partial_i \psi-\delta^i_j \, \partial_k \psi \right)|_0\\
&-\frac{1}{6} \, x^i\, x^j\, x^l \left(\delta^i_k \,\partial_{jl} \psi+\delta^j_k\, \partial_{il} \psi-\delta^i_j \, \partial_{ki} \psi \right)|_0
\end{split}
\ee
There is still the freedom to choose the value of the coordinates of the central world-line at the initial proper-time $\tau_i$; one can set
\be
p^i(\tau_{i})=0\,,\qquad p^i(\tau)=\int_{\tau_i}^\tau v^i(\tau', 0) \,
d\tau'\, .
\ee
Focusing on the scalar sector we can always write the 3-velocity as a gradient
\be
\begin{split}
\int_{\tau_i}^\tau v^k|_p d\tau & = \partial_k V(|\textbf{x}|)|_{x^i=p^i}
\equiv x^k {\cal V}(x)\,.
\end{split}
\ee
It is worth stressing that it is necessary to consider the
 relation between CFC and comoving coordinates at least at third order
to get the correct off-diagonal spatial metric corrections at second order in $x^i$, which have the form $x^i x^j\partial_{ij} h$. \\
In order to find the function $\lambda$ which defines the deformed
dilatation one has to express $\psi|_0$, $\partial_i\psi|_0$ and
$\partial_{ij}\psi|_0$ as a Taylor series centred on x, at least at
third order  in $x^i$. After some tedious computation one gets, from  eq. (\ref{CFC_form}):
\be
 x_F^k=\left[1+\zeta-\frac{1}{2} |\textbf{x}|\, \partial_x \zeta(|\textbf{x}|)+\frac{1}{6} |\textbf{x}|^2\, \partial_x^2 \zeta(|\textbf{x}|)-\frac{1}{12} |\textbf{x}|^3\,\partial_x^3 \zeta(|\textbf{x}|)- {\cal V}(|\textbf{x}|)\right]x^k\,,
\ee
where we have  imposed $\psi(x)=\psi(|\textbf{x}|)$.
Thus, the transformation between spatial CFC and spatial comoving
coordinates perfectly matches a space-time dependent dilatation with
\be
\lambda=\zeta-\frac{1}{2} |\textbf{x}|\, \partial_x \zeta(|\textbf{x}|)+\frac{1}{6} |\textbf{x}|^2\, \partial_x^2 \zeta(|\textbf{x}|)-\frac{1}{12} |\textbf{x}|^3\,\partial_x^3 \zeta(|\textbf{x}|)- {\cal V}(|\textbf{x}|)+O(|\textbf{x}|^4)\,.
\ee
This simple procedure can be generalized at any order
\be  
\label{Lambda_CFC}
\lambda^{(n)}=\sum_{l=0}^n \,\alpha^{(l)} \, |\textbf{x}|^l\,
\partial_x^l \zeta(|\textbf{x}|) - {\cal
  V}(|\textbf{x}|)+O(|\textbf{x}|^{n+1})\,, \qquad \alpha^{(0)}=1 \, .
\ee
Of course, as discussed in the main text, $\zeta$ is the
gauge-invariant $\zeta$.  Once a
central world-line is chosen as the origin of the new {\it Lagrangian}
coordinate system, on can take $x$ to be arbitrarily (not necessarily
small) by  truncating the series to a sufficiently large $n$-th
order. Finally, even if $x$ is taken to be  a small displacement from
the  central value (set to zero), in Fourier space any information
about the smallness of a term $x^i \partial_i$ is lost for any value
assumed by $x$, as shown  in Section \ref{grad_exp}.
If we neglect the ${\cal V}$ presence, the new $\tilde g_{ij}$ take the following form
\be
\label{CFC_metric}
 \tilde g_{ij}=\frac{1}{3}\,x^k \,x^l\, \left(\delta^l_j \, \partial_{ik}+\delta^l_i \, \partial_{jk} -\delta^i_j\, \partial_{lk}-\delta^k_l\, \partial_{ij}\right)\psi+O(x^3)\,,
\ee
which is the same result obtained in \cite{Dai:2015jaa}. Still in the
series of works
\cite{Pajer:2013ana}\cite{Dai:2015jaa}\cite{Dai:2015rda}\cite{Cabass:2016cgp}\cite{Cabass:2018roz}
it is assumed that the CFC transformation works
in a region on an  {\it unspecified} scale, depending on the dimension of the patch of Universe under investigation. Thus, all the quantities which appear in the previous relations are coarse-grained. In particular, this holds for eq. (\ref{Lambda_CFC}), where an Heaviside theta window function, defined as in (\ref{window}), is understood. Finally, we stress once again that even if in eq. (\ref{CFC_metric}) the diagonal part of the spatial metric proportional to $\psi \delta_{ij}$ disappears, this does not mean that the $\psi$ function is shifted.

\bibliographystyle{unsrt}  
\bibliography{paper}

\end{document}